\documentclass[a4paper]{article}

\usepackage{17lomcon}        
\usepackage{cite}             
\usepackage{epsfig}           

\begin{document}


\title{PROMPT NEUTRINO FLUX IN THE ATMOSPHERE REVISITED}

\author{Maria Vittoria Garzelli~\footnote{Invited contribution to the Proc. of the 17$^{th}$ Lomonosov Conference on Elementary Particle Physics, August 20 - 26 2015, Moscow State University, Moscow, Russia}, 
        Sven-Olaf Moch, 
        and
        G{\"u}nter Sigl 
}

\affiliation{
II Institute for Theoretical Physics, University of Hamburg, \\
Luruper Chaussee 149, D-22761 Hamburg, Germany 
}


\date{}
\maketitle


\begin{abstract}
Prompt neutrino fluxes due to the interactions of high-energy cosmic rays
with the Earth's atmosphere are backgrounds in the search for high-energy
neutrinos of galactic or extra-galactic origin 
performed by Very Large Volume Neutrino Telescopes. 
We summarize our predictions for prompt neutrinos, showing their
basic features as emerging from the calculation in a QCD framework
capable of describing recent charm data from the Large Hadron Collider. 
\end{abstract}

\section{Motivation of this work}
Very large volume neutrino telescopes (VLV$\nu$Ts) provide a unique window
to the high-energy universe~\cite{Kappes:2015woa}, because they allow for the detection of high-energy neutrinos which can reach the Earth after having travelled from far galactic or extra-galactic sources, almost undeflected by cosmic magnetic fields and insensitive to electromagnetic and strong forces.
In this respect neutrinos propagate quite differently with respect to 
charged cosmic rays (CRs), and are thus precious messengers capable of providing direct information on the regions of the Universe where they are produced.

A first detection of high-energy leptonic events up to an energy of a few PeV
was reported by the IceCube collaboration. The initial evidence 
from the first two-year data analysis~\cite{Aartsen:2013jdh} was subsequently strenghtened by increased statistics in the three and four-year analyses~\cite{Aartsen:2014gkd, Montaruli:2015six}. So far, more than 50 high-energy starting events have been recorded. The discussion on the origin of these events is quite lively, and a number of hypotheses concerning the possible sources has been formulated. Although it is believed that most of the neutrino events detected so far have an extraterrestrial origin, theoretical calculations predict the existence of a terrestrial component due to the decays of the unstable hadrons produced by the interactions of high-energy and ultra-high-energy CRs with the Earth's atmosphere~\cite{Gondolo:1995fq}.
This terrestrial component should be properly taken into account as a background and subtracted in any astrophysical a\-na\-ly\-sis. The terrestrial component
is usually separated into two contributions: the conventional neutrino flux,
coming from the decay of $\pi^\pm$ and $K^\pm$ mesons, 
and the prompt neutrino flux,
originating from the decay of heavier mesons and hadrons, in particular those including charm quarks as valence quarks in their composition.
The prompt component is supposed to become dominant with respect
to the conventional one only at energies high enough.   
Several computations of conventional neutrino fluxes exist~(see e.g. Ref.~\cite{Honda:2006qj}), together
with experimental evidence. 
On the other hand, no experimental evidence for the existence of a prompt flux has emerged so far. It is thus important to provide precise and  
refined theoretical predictions
of prompt neutrino fluxes, to understand if the lack of  
detection is just due to the still low statistics of the VLV$\nu$T experiments at high energy, or if the methodologies and/or theories used so far to predict these fluxes have~some~flaws. 

\section{Charm hadroproduction in QCD and prompt neutrino fluxes}

In Ref.~\cite{Garzelli:2015psa} we have recently computed prompt neutrino fluxes making use of QCD, with microscopic interactions described by event generators including up-to-date information from the QCD theory community working at hadron col\-li\-ders. As shown in the following, we have used the same generators adopted in our astrophysical study, to produce predictions which can be compared with experimental data recently collected at the Large Hadron Collider (LHC).

Nowadays, state-of-the-art tools for predicting many differential distributions at the LHC are represented by Shower Monte Carlo (SMC) codes, matched with hard-scattering processes evaluated in perturbative QCD including next-to-leading-order (NLO) radiative corrections at least~\cite{Nason:2012pr}. For a few selected processes, calculations at the differential level involving the resummation of different kinds of large logarithms to all orders, or next-to-NLO (NNLO) QCD radiative corrections, start to be available as well. However, this is not yet the case for the processes involving charm quarks we are interested in in this paper. Interfacing  hard scattering amplitudes including radiative corrections with the parton shower algorithms (PS) embedded in the SMCs requires a careful matching, to avoid double counting effects and to ensure a proper coverage of the whole phase space available for light parton emissions. At present, one of the most widely used NLO QCD + PS matching schemes is POWHEG~\cite{Nason:2004rx, Frixione:2007vw}, 
which has been automated in dedicated tools, like e.g. the \texttt{POWHEGBOX}~\cite{Alioli:2010xd}. 
At NNLO the situation is even more complicated, and efforts to extend matching schemes with PS at this level of accuracy are ongoing, as well as those for pushing the accuracy of PS far beyond the present one.

In this paper we consider $c\bar{c}$ $\rightarrow$ $c$-hadron + $X$ hadroproduction evaluated by means of \texttt{POWHEGBOX}~+~\texttt{PYTHIA 6}~\cite{Sjostrand:2006za}. This allows to generate differential distributions for the hadroproduction of different kinds of charmed mesons and hadrons at NLO QCD + PS accuracy. The PS ensures at least leading logarithmic (LL) accuracy in all collinear QCD emissions following the first light partonic one accompanying $c\bar{c}$,
which is instead evaluated by \texttt{POWHEGBOX} with NLO accuracy. 

\begin{figure}[t!]
\begin{center}
\includegraphics[width=0.44\textwidth]{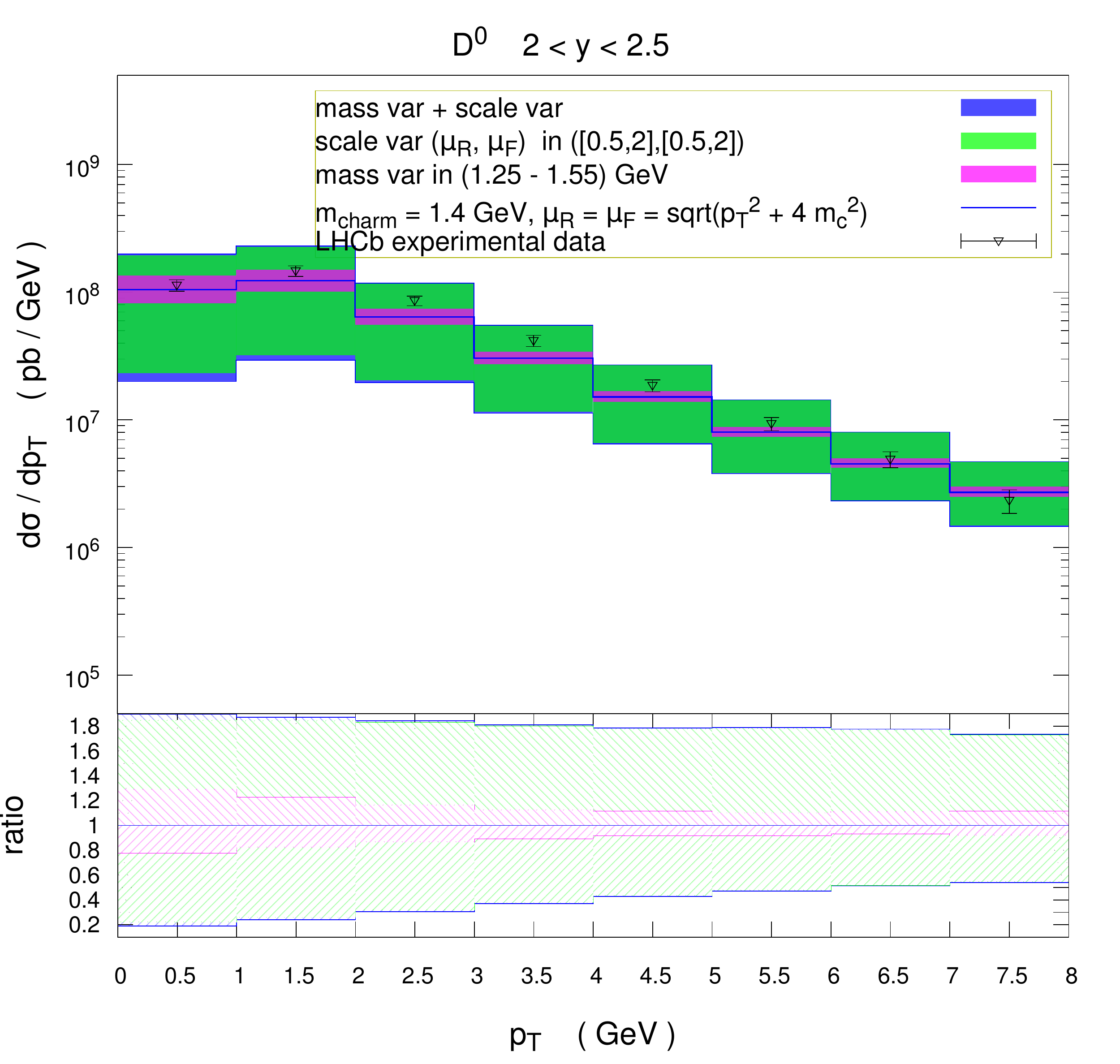}
\includegraphics[width=0.44\textwidth]{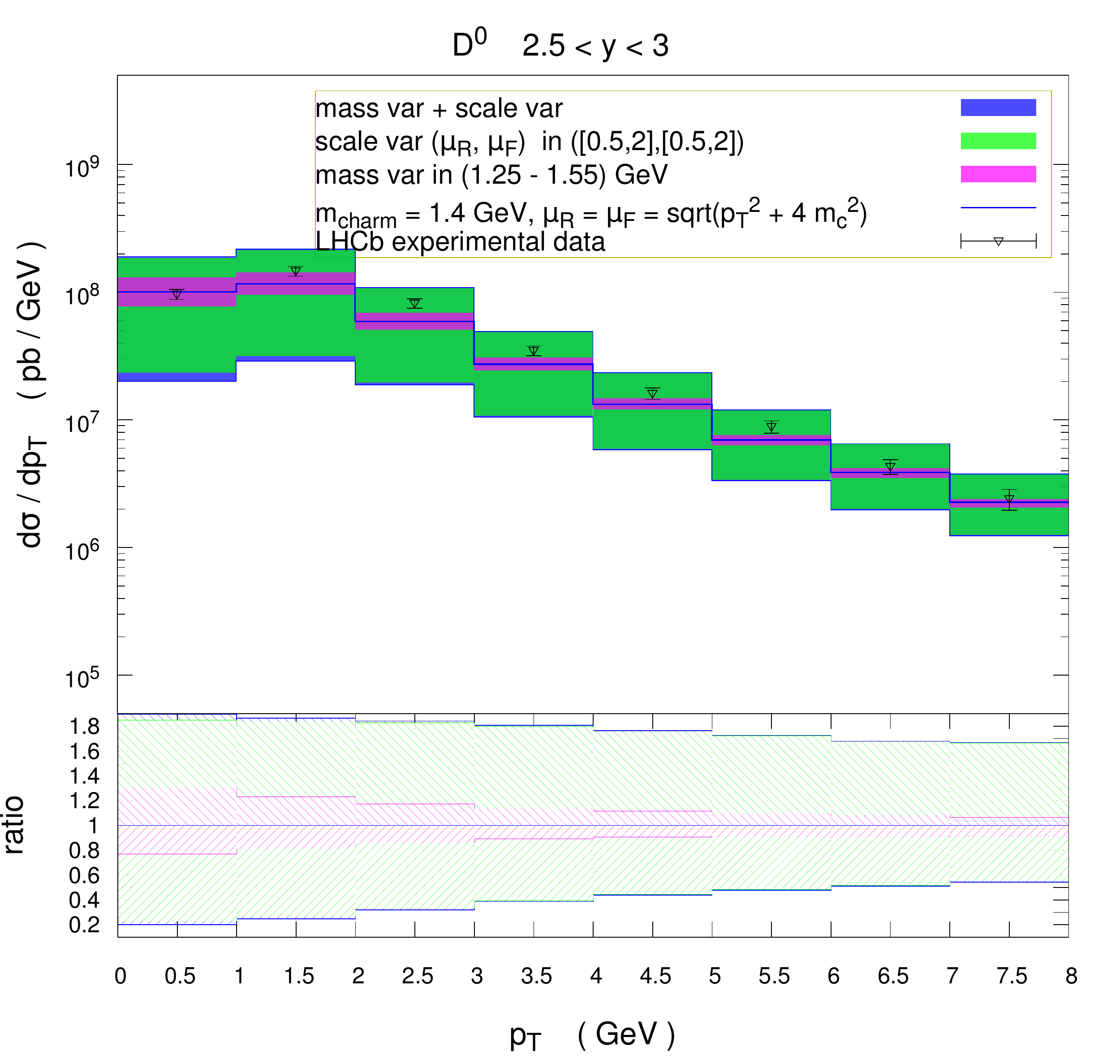}
\includegraphics[width=0.44\textwidth]{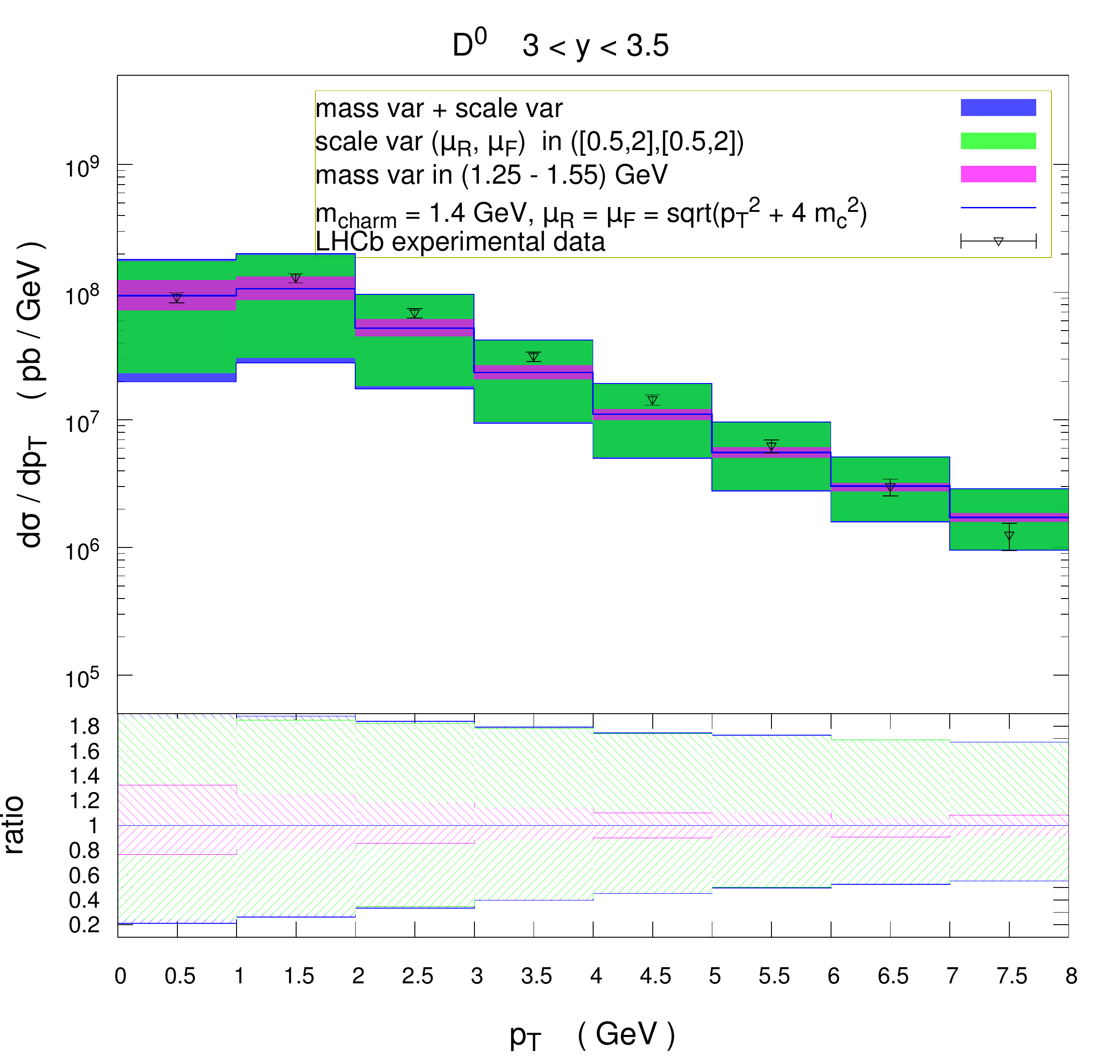}
\includegraphics[width=0.44\textwidth]{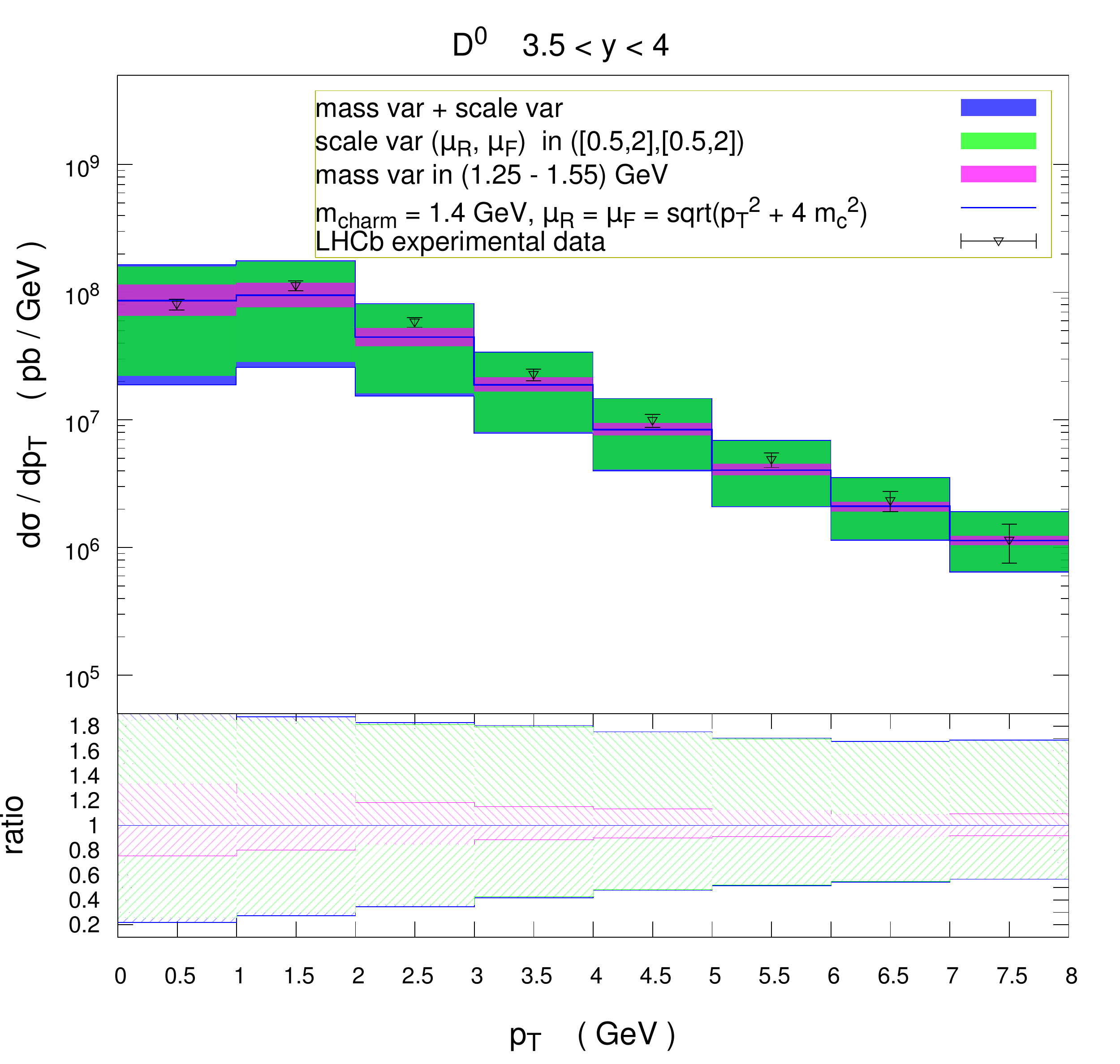}
\includegraphics[width=0.44\textwidth]{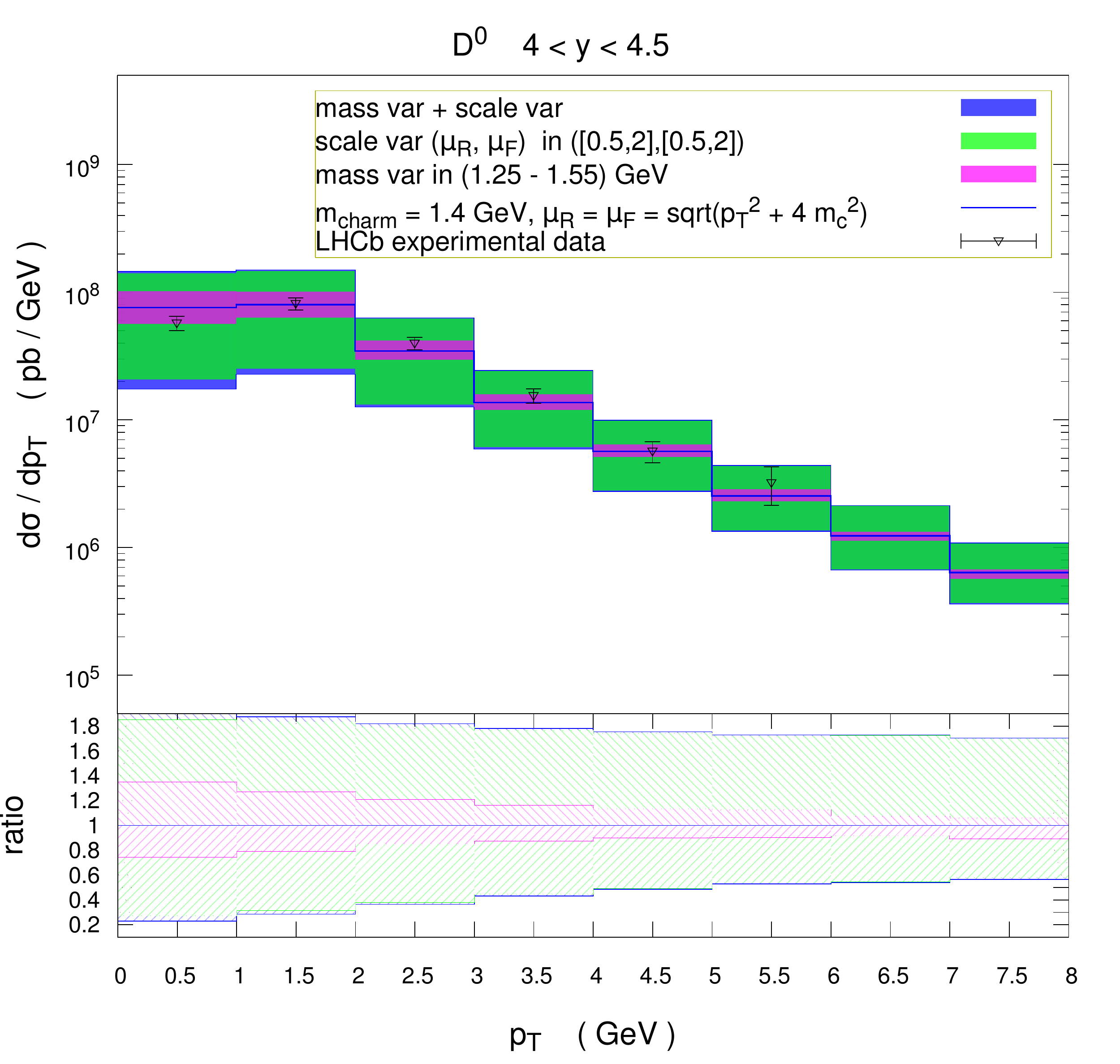}
\end{center}
\caption{\label{rapidity} Theoretical predictions for $D^0$ hadroproduction by {\texttt{POWHEGBOX~+~PYTHIA 6}} compared to LHCb data at 7 TeV. Transverse momentum distributions in different rapidity bins are shown in different panels. The green and violet bands refer to scale and mass variation, respectively, summed in quadrature in the blue band. Central values for all input parameters are fixed as explained in the text. 
}
\end{figure}

This computation requires to fix some input. To establish the central va\-lues of the renormalization and factorization scales $\mu_R$ and $\mu_F$, of the charm mass $m_c$ and a set of parton distribution functions (PDFs) suitable for the calculation of prompt neutrino fluxes, taking into account that a broad range of center-of-mass energies $\sqrt{s}$ are involved, including both energies much lower and much higher than those explored so far at colliders, we have studied the behaviour of the total $c\bar{c}$ hadroproduction cross-section at different orders/degrees of accuracy. We have compared predictions including LO, NLO and NNLO QCD corrections, as well as the behaviour of the cross-sections when adopting  the $\overline{MS}$ scheme as an alternative to the on-shell one in the renormalization of the charm quark mass. The procedure leading to our choice of the central values of these parameters and of their range of variation, useful to determine the associated uncertainties, is detailed in Ref.~\cite{Garzelli:2015psa}. Here we summarize our choices. The central $\mu_R$ and $\mu_F$ scales were fixed to $\mu_0$=$\sqrt{p_{T,c}^2 + 4 m_c^2}$, and allowed to vary in the range ($\mu_R$, $\mu_F$) $\in$ (0.5, 0.5), (2,2), (0.5, 1), (1, 0.5), (1, 2), (2,1) $\mu_0$. We did not include the variations (0.5, 2) and (2, 0.5) $\mu_0$, as suggested in Ref.~\cite{Cacciari:2012ny}. The central charm mass value was fixed to $m_{c}$ = 1.4 GeV in the on-shell scheme and allowed to vary in the range $\pm$~0.15~GeV. We have used the ABM11 PDFs~\cite{Alekhin:2012ig} at NLO including the PDF and  $\alpha_s$ variations.  
We emphasize that the choice of all these parameters was uniquely derived by theoretical considerations, taking into account in particular the convergence of the perturbative series of radiative corrections, and the behaviour of different sets of PDFs already available in the {\texttt{LHAPDF 6.1.5}} interface~\cite{Buckley:2014ana} in an extended Bjorken-$x$ range, also covering the region of low $x$. No comparison with any recent experimental data was used to tune these parameters or retune other parameters entering {\texttt{PYTHIA 6}}, for which we have adopted one of the most recent already available Perugia tunes~\cite{Skands:2010ak}, ensuring a transverse-momentum ordered PS. 

\begin{figure}[t!]
\begin{center}
\includegraphics[width=0.44\textwidth]{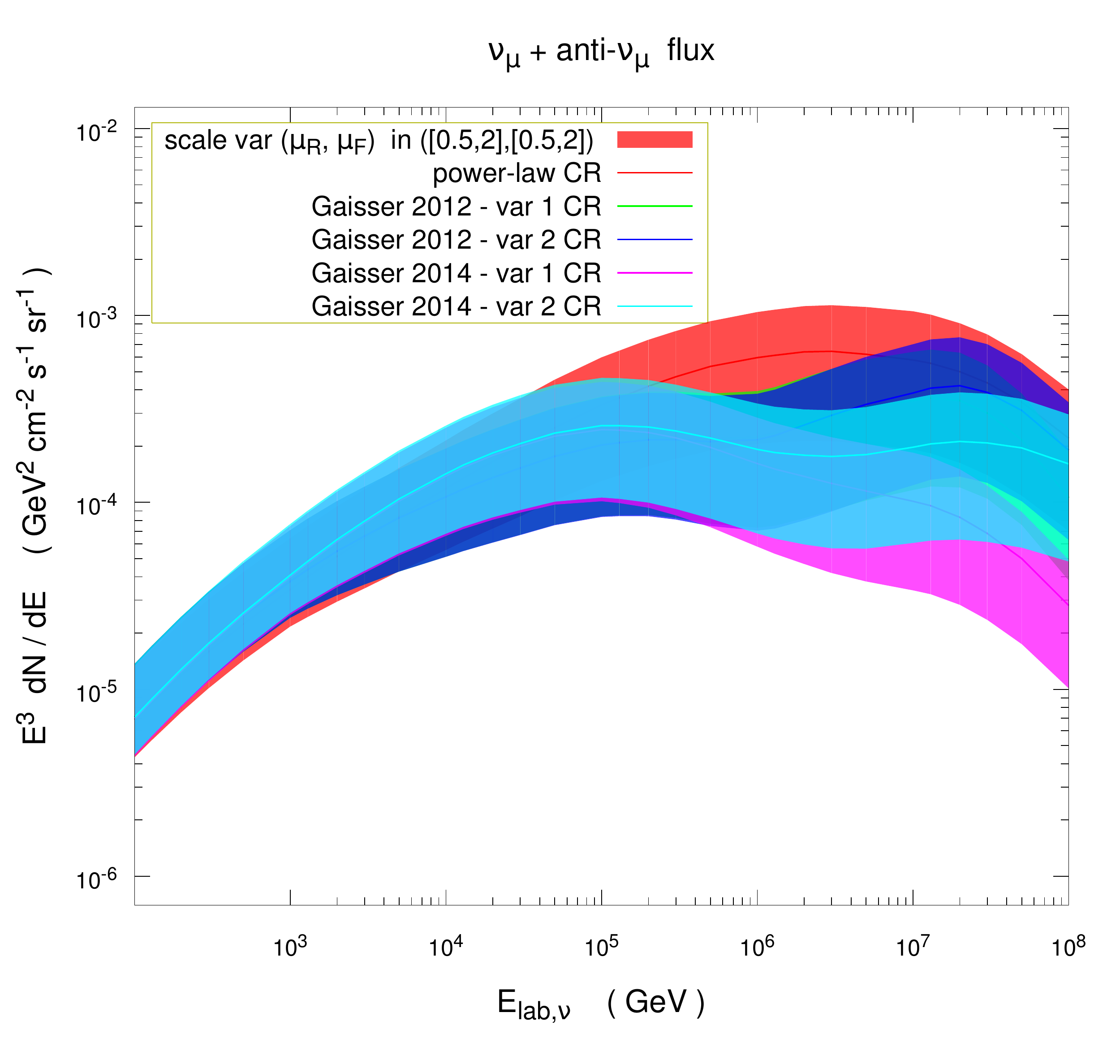}
\includegraphics[width=0.44\textwidth]{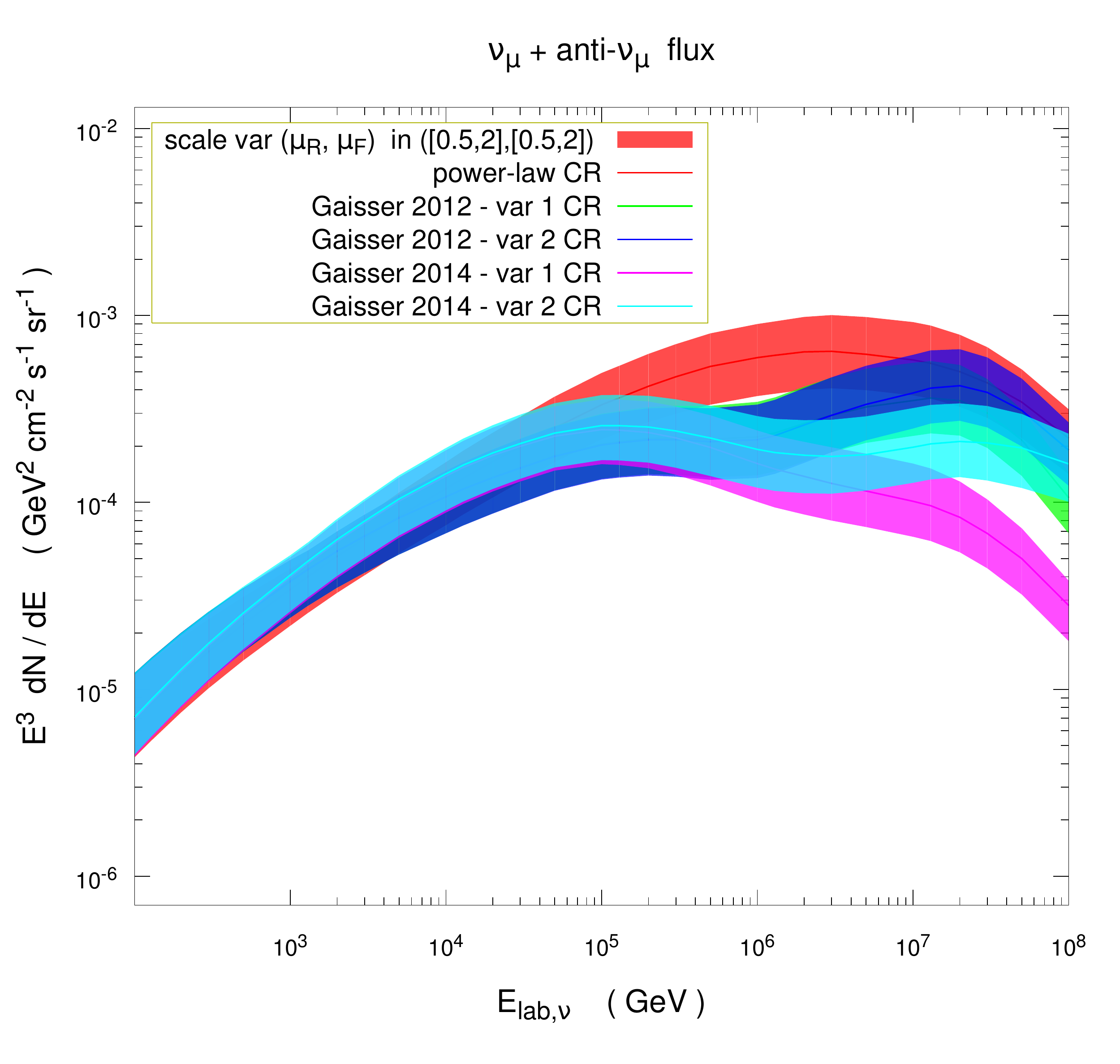}
\includegraphics[width=0.44\textwidth]{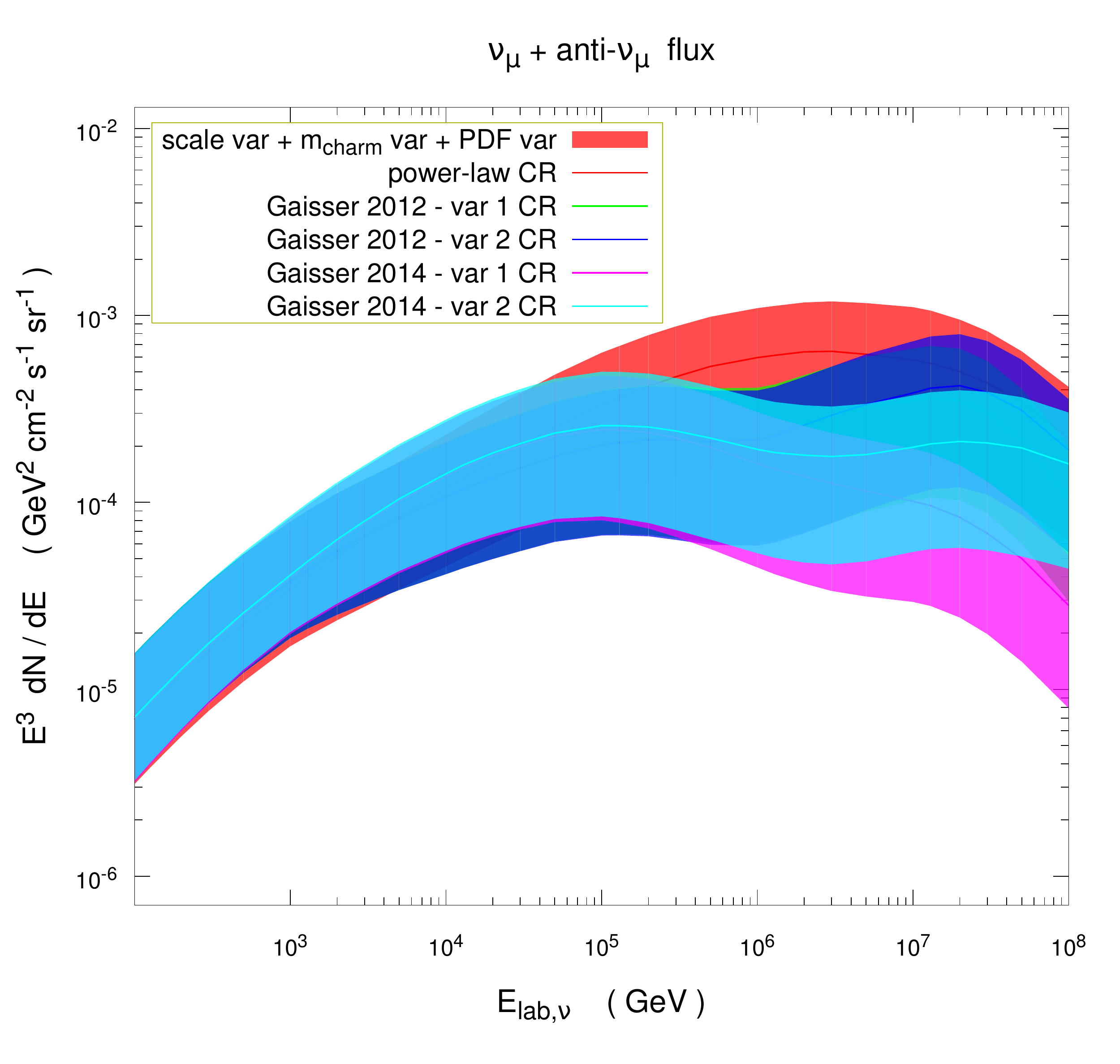}
\includegraphics[width=0.44\textwidth]{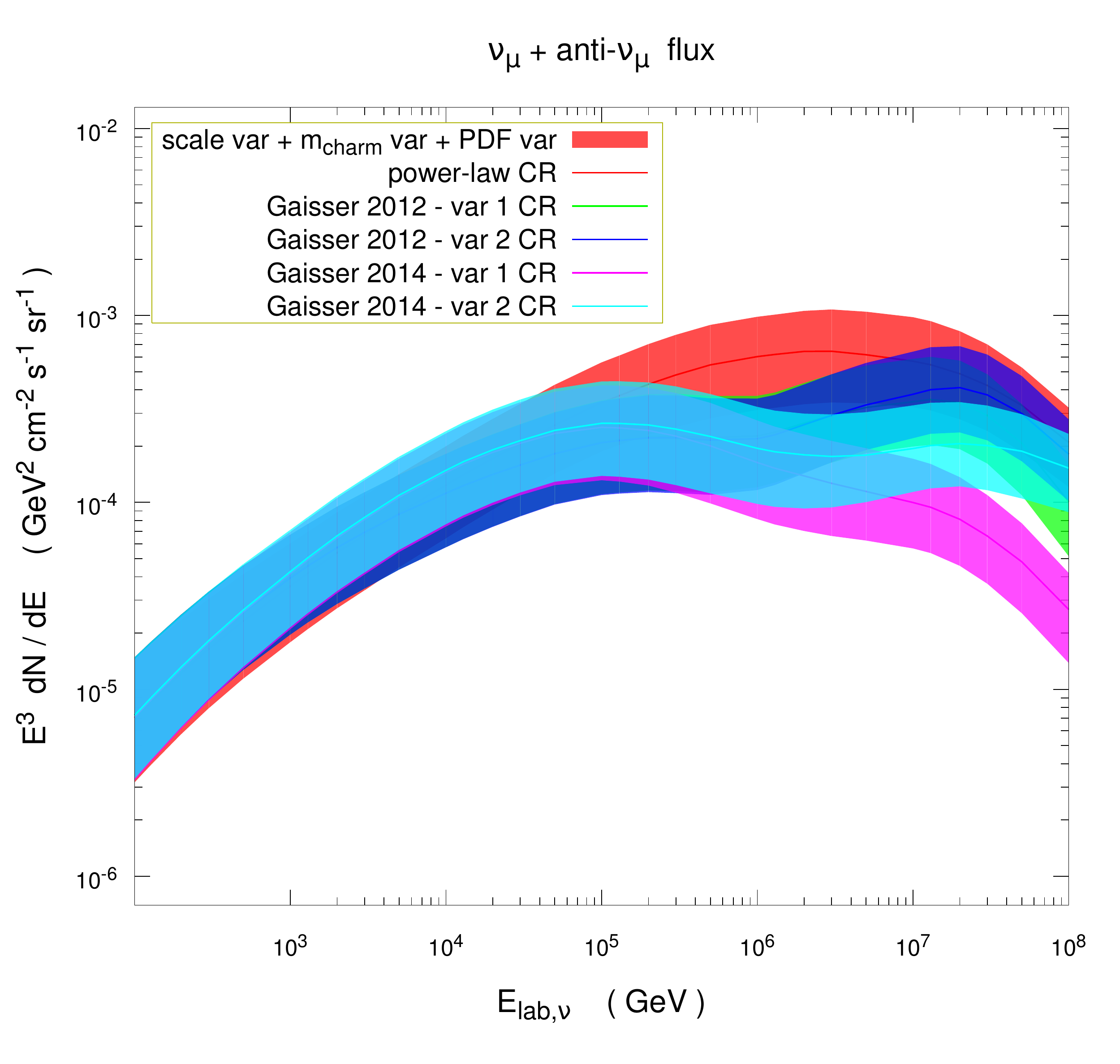}
\end{center}
\caption{\label{fluxes} Our predictions for ($\nu_{\mu}$ + $\bar{\nu}_\mu$) fluxes as a function of $E_\nu$. Uncertainties due to scale variation around the central value $\mu_R$ = $\mu_F$ = $\sqrt{p_T^2 + 4 m_c^2}$, are shown in panel 2.a (upper-left), considering all possible ($\mu_R$, $\mu_F$) scale combinations (0.5, 0.5), (2, 2), (1, 2), (2,~1), (0.5,~1) and (1, 0.5) $\mu_0$, and in panel 2.b (upper-right) by disregarding the non-diagonal contributions (0.5,~1) and (1,~0.5) $\mu_0$. The addition in quadrature of the uncertainties in panel 2.a (2.b) with those due to $m_c$ and PDF variation gives rise to the QCD uncertainty 
band in panel 2.c (2.d).  In each panel, different colors refer to fluxes computed with different primary CR all-nucleon spectra, including power-law and four different variants of Gaisser CR spectra~\cite{Gaisser:2013bla, Stanev:2014mla}. See text for more detail.}
\end{figure}

Comparison with experimental data was done a-posteriori, to test the theoretical predictions. In particular the LHCb collaboration has recently measured the hadroproduction of charmed mesons and baryons in different rapidity and transverse momentum bins, at both 7 and 13 TeV~\cite{Aaij:2013mga, Aaij:2015bpa}, covering the total range 2~$<$~$y$~$<$ 4.5, corresponding to mid-peripheral collisions. A comparison between our theoretical predictions for $D^0$ hadroproduction and experimental data at 7 TeV is shown in Fig.~\ref{rapidity}. The theoretical predictions turn out to agree with experimental data within the theoretical uncertainty bands due to scale and mass variation in all rapidity bins. Similar agreement is found for charged $D$-mesons.

The large uncertainties due to $\mu_R$ and $\mu_F$ scale variation, shown in Fig.~\ref{rapidity} 
for LHC predictions, are almost independent of $\sqrt{s}$, i.e. remain of the same order of magnitude even at other energies, and propagate to prompt neutrino fluxes. Therefore, they can be considered to be the dominant QCD uncertainty, as follows from the comparison of Fig.~\ref{fluxes}.a 
with with Fig.~\ref{fluxes}.c. 
In particular, the non-diagonal $\mu_R$ and $\mu_F$ variations, neglected in calculations completed before our one, like e.g. that in Ref.~\cite{Bhattacharya:2015jpa}, turn out to enlarge the uncertainty band on prompt neutrino fluxes of several tens percent, as shown in Fig.~\ref{fluxes} (compare Fig 2.a with Fig. 2.b or Fig 2.c with Fig 2.d). 

At the highest energies, uncertainties related to our poor knowledge of the composition of the primary flux of CRs entering the upper layer of the Earth's atmosphere, turn out to become even larger than those of QCD origin, as shown in Fig.~\ref{fluxes}, and dedicated efforts from extended air shower (EAS) experiments are indeed needed to reduce them, together with the improvement and reduction of uncertainties in the theoretical models describing EAS formation. 
\section*{Acknowledgments}
We are grateful to the Conference organizers and partecipants for many
lively discussions. 
This work has been supported by Deutsche 
Forschungsgemeinschaft in Sonderforschungsbereich 676.



\end{document}